\documentclass[journal=jctcce,manuscript=article]{achemso}

\usepackage[version=3]{mhchem} % Formula subscripts using \ce{}
\usepackage{amsmath}
\usepackage{caption}
\usepackage{subfig}
\usepackage{graphicx}
\usepackage{epstopdf}

\newcommand{\tn}{\textnormal}
\author{Kai Wang}
\author{Shiyang Long}
\author{Zhiming Zhang}
\author{Lanru Liu}
\author{Qimeng Wang}
\author{Pu Tian}
\email{tianpu@jlu.edu.cn}
\phone{+86-431-85155287}
\affiliation
{College of Life Science}
\affiliation
{College of Life Science}
\affiliation
{College of Life Science}
\affiliation
{College of Life Science}
\affiliation
{College of Life Science}
\alsoaffiliation
{MOE Key Laboratory of Molecular Enzymology and Engineering\\
Jilin University \\
2699 Qianjin Street, Changchun 130012}

\title[]{Ideal gas behavior of rotamerically defined conformers in native globular proteins}

\abbreviations{}
\begin{document}

%%%%%%%%%%%%%%%%%%%%%%%%%%%%%%%%%%%%%%%%%%%%%%%%%%%%%%%%%%%%%%%%%%%%%
%% The manuscript does not need to include \maketitle, which is
%% executed automatically.  The document should begin with an
%% abstract, if appropriate.  If one is given and should not be, the
%% contents will be gobbled.
%%%%%%%%%%%%%%%%%%%%%%%%%%%%%%%%%%%%%%%%%%%%%%%%%%%%%%%%%%%%%%%%%%%%%
\begin{abstract}
Protein side chain entropy has been found to be important by both experimental and computational studies. However, the connection between side chain torsional states and protein conformational distributions remains vague. Based on the robustness of side chain rotameric states observed in both experimental structures and large scale molecular dynamics simulations, we propose to define unique combinations of side chain rotameric states as basic conformers, termed RCONFs, for entropy calculation. 
%As the number of RCONFs is astronomically large for even a small protein of less than 100 residues, 
we further hypothesize that all RCONFs have the same constant local configurational integral for a given protein under specified solvent conditions. It follows from this hypothesis that RCONFs behave like ideal gas configurations, RCONF based conformational entropy may be effectively expressed as $S = lnW$, with $W$ being the number of RCONFs that are thermally accessible, and change of free energy between two given macrostates is equivalent to that of conformational entropy with a mere difference of a negative temperature factor. The validity of the ideal gas hypothesis and inferred property of change of both conformational entropy and free energy is tested in extensive molecular dynamics (MD) simulation trajectories of six globular proteins in native state. The advantage of the corresponding end point free energy method is discussed.
\end{abstract}

\section*{Introduction}
While theoretical importance of entropy in physical systems has been appreciated for a long time\cite{Jaynes1957,Jaynes1957b,Wehrl1978}, experimental evidence for decisive roles of conformational entropy in bimolecular interactions appeared only recently\cite{Frederick2007,Tzeng2012,Kasinath2013,Wand201375}. This is due to the fact that decomposition of free energy into contributions from system comprising components and their correlations is extremely challenging. The total free energy change of a typical protein-ligand system may be written as:
\begin{align}
\Delta G &=  \Delta H - T\Delta S \\
\Delta H &= \Delta H_{pp} + \Delta H_{ll} + \Delta H_{vv} + \Delta H_{pl} + \Delta H_{lv} + \Delta H_{pv} \\
\Delta S &= \Delta S_{pp} +\Delta S_{ll} + \Delta S_{vv} + \Delta S_{cross}
\end{align}
With letters $p$, $l$, and $v$ in subscript represent protein, ligand and solvent respectively. Repeated subscripts ($pp$, $ll$ and $vv$) represent molecular interactions in the same type of molecules in enthalpic terms and corresponding entropy in entropic terms. Mixed subscripts ($pv$, $pl$ and $lv$) represent molecular interactions between different type of molecules in enthalpic terms. $\Delta S_{cross}$ is the change of entropy reduction due to correlation of molecular degrees of freedom (DOFs) between (among) different types of molecules in entropic terms\footnote{$S_{pp}$ is the entropy of protein molecules under the free energy landscape determined by both the intra- and inter-molecular interactions existing under given solvent conditions. $S_{ll}$ and $S_{vv}$ are defined similarly. When ligands are unbound, $S_{pl} \approx 0$ in dilute solution and $S_{cross} = S_{pv} + S_{lv}$. When ligands are bound, there is actually a third order correlation term $S_{plv}$ so $S_{cross} = S_{pl} + S_{pv} + S_{lv} + S_{plv}$.}.
%Detailed free energy decomposition into terms listed in the above equations (2) and (3), especially calculation of correlation entropy terms $\Delta S_{pv}, \Delta S_{lv} and \Delta S_{pl}$, is extremely challenging. 
Computationally, systematic and quantitative evaluation of numerous methods for direct calculation of macromolecular entropy and solvent entropy is yet to be done\cite{Reinhard2007,Tyka2007,Reinhard2009,Wang2009,Gerogiokas2014,Suarez2014}. Furthermore, no general method is available for direct computation of correlation entropies between solute and solvent molecules (e.g. protein and water). 
%We are primarily interested in the entropy change of proteins ($\Delta S_{pp}$). If we consider limit of dilute solution and neglect change of rotational and translational entropy, then $\Delta S_{pp}$ is reduced to change of configurational entropy $\Delta S_{config}$. 
Experimentally, the well-established isothermal calorimetry (ITC)\cite{Falconer2010} measurement only directly gives $\Delta H$ and $\Delta G$. The ``model free approach'' of Lipari and Szabo\cite{Szabo1982,Szabo1982b} provides a theoretical connection between local motion of bond vectors, which is measurable by NMR relaxation and characterizable by the square of the generalized order parameter, and thermodynamics. In a series of NMR studies\cite{Kasinath2013,Wand201375}, change of protein configurational (conformational) entropy as manifested by side chain motion of methyl bearing residues was found to exhibit strong linear correlation with total entropy change upon binding of ligands. Consequently, the ensemble average of side chain entropies of methyl bearing residues may effectively serve as an ``entropy meter'' for binding entropy. This strategy was successfully utilized by Tzeng \emph{et. al.}\cite{Tzeng2012} in elegantly designed CAP (catabolite activator protein) -DNA interaction systems to demonstrate the importance of configurational (conformational) entropy contribution to the free energy. Despite these significant advances, the connection between protein side chain rotameric states and thermodynamics of protein conformational distributions remains to be constructed.

%However, measuring entropies of methyl bearing residues is ( comparable with, ) or more expensive than, that of calorimetric measurements of binding entropy in terms of cost. Additionally, correlation coefficients for arbitrarily given proteins are not easily predictable. Therefore, quantitative application of this great discovery for prediction of bimolecular interactions in its present form is severely limited even if we assume that side chain entropies of methyl bearing residues may be reliably predicted from molecular dynamics simulations as demonstrated by one of the study.
\section*{The ideal gas hypothesis of rotamerically defined conformers}
For a protein molecule, a binding event with a ligand is essentially a redistribution of equilibrium statistical weight in the conformational space. The observed linear correlation between entropy change of methyl bearing side chains and that of ligand binding suggests that there might be undiscovered general rule(s) governing distributions of rotamerically defined conformers (RCONFs) in the conformational space.  To be more specific on the definition of RCONFs, each RCONF is defined by a given combination of torsional states of unique side chain all-heavy-atom torsional DOFs. Two structural states of a protein belong to the same RCONF if and only if they share the same torsional state for each unique side chain all-heavy-atom torsional DOF. 

The configurational integral of a protein molecule in solution may be written as:
\begin{equation}
Z = \int e^{-\beta U(r_p, r_v)}dr_pdr_v
\label{config-int}
\end{equation}
with $\beta$ being the reciprocal temperature, $U$ being the potential energy, and $r_p$ and $r_v$ being coordinates of protein and solvent atoms. If we partition the whole conformational space into small basic elements, each of which corresponds to a unique RCONF, the configurational integral can be written as a sum of all sub-integrals (local integrals) corresponding to RCONFs:
\begin{align}
Z  & = \sum^{N_{rconf}(U(r_p, r_v))}_{i=1}Z_i \\
Z_i & = \int_{rconf_i}e^{-\beta U(r_p, r_v)}dr_pdr_v
\end{align}
For two arbitrarily given macrostates $A$ and $B$:
\begin{align}
Z^A & = \sum^{N^A_{rconf}(U(r_p, r_v))}_{i=1}Z^A_i \\
Z^B & = \sum^{N^B_{rconf}(U(r_p, r_v))}_{j=1}Z^B_j  \\
Z^A_i & = \int_{rconf^A_i}e^{-\beta U(r_p, r_v)}dr_pdr_v \\
Z^B_j & = \int_{rconf^B_j}e^{-\beta U(r_p, r_v)}dr_pdr_v \\
\Delta F^{AB} & =  k_BTln\frac{Z^A}{Z^B}
\end{align}
$\Delta F^{AB}$ is the change of Helmholtz free energy between macrostates $A$ and $B$. 
If we consider limit of dilute solution and neglect change of rotational and translational entropy, then $\Delta S_{pp}$ in equation (3) is reduced to change of configurational entropy $\Delta S_{config}$. Based on the definition of RCONFs and following a common practice in theoretical studies\cite{Karplus1987,Chang2004,Chang2007,Numata2012}, we may split configurational entropy $S_{config}$ into conformational and vibrational contributions (a physics based proof is provide by Chang and Gilson\cite{Chang2007}) as shown below: %with the relative importance of these two terms depending upon specific definition of the conformation.

\begin{align}
S_{config} &=  S_{rconf}+ S_{rconf\tn{-}vib} \\
S_{rconf} & = -k_B\sum^{i=N_{rconf}(U(r_p, r_v))}_{i=1}P_i lnP_i \\
S_{rconf\tn{-}vib} & = \sum^{i = N_{rconf}(U(r_p, r_v))}_{i=1}P_iS^i_{rconf\tn{-}vib} 
\end{align}
For two arbitrarily given macrostates (conformation) $A$ and $B$,
\begin{align}
S^A_{config} & = S^A_{rconf} + S^A_{rconf\tn{-}vib} \\
S^A_{rconf} & = -k_B\sum^{i=N^A_{rconf}(U(r_p, r_v))}_{i=1}P^A_i lnP^A_i \\
S^A_{rconf\tn{-}vib} & = \sum^{i = N^A_{rconf}(U(r_p, r_v))}_{i=1}P^A_iS^{A_i}_{rconf\tn{-}vib}  \\
S^B_{config} & = S^B_{rconf} + S^B_{rconf\tn{-}vib} \\
S^B_{rconf} & = -k_B\sum^{j=N^B_{rconf}(U(r_p, r_v))}_{j=1}P^B_j lnP^B_j \\
S^B_{rconf\tn{-}vib} & = \sum^{j = N^B_{rconf}(U(r_p, r_v))}_{j=1}P^B_jS^{B_j}_{rconf\tn{-}vib}  \\
\Delta S^{AB}_{config} & = \Delta S^{AB}_{rconf} + \Delta S^{AB}_{rconf\tn{-}vib} 
\end{align}
$S_{rconf}$ is the conformational entropy based on our definition of RCONFs, $S_{rconf\tn{-}vib}$ include contributions from both bonding and bending vibrational entropies, and from local part of torsional entropy within each RCONF, and $P^{A(B)}_{i(j)}$ is the probability of the $i(j)$th RCONF in the macrostate $A(B)$. The rational of only considering side chain torsional DOFs will be discussed after presentation of the main results. As is apparent from equations (12-14), allocation of $S_{config}$ between the ``conformational term'' and the ``vibrational term'' depends on specific definition of conformations. The robustness of rotameric states as observed in both experimental structures\cite{Bower1997,Shapovalov2011} and MD simulations\cite{Scouras2011} provides a feasible and practical base for general applicability of RCONF-based conformational entropy. Before proceeding further, we need to clarify that two terms ``configurational entropy'' and ``conformational entropy'' are used interchangeably in many experimental studies\cite{Tzeng2012,Kasinath2013,Wand201375}. In this study, the term $S_{rconf}$ denotes the entropy based on distributions of RCONFs, 
%and the term $S_{confor}$ denotes the entropy based on distributions of conformations defined in other ways. 
it does not consider details of microstate distributions within any given RCONF, and is apparently different from $S_{config}$, which includes all vibrational contributions.

%Since it is demonstrated that entropic contributions of hard DOFs (bonding and bending) is separable from that of soft ( torsional ) DOFs\cite{Li2009}, and our primary goal is to investigate how RCONFs distribute in the protein conformational space, we leave out the complexity of analyzing bonding and bending vibrational entropies in this study.

For typically-sized natural proteins, $N_{rconf}$ is an astronomically large number. For example, with fixed backbone and rigid rotamer model, the allowed combination of side chain rotameric states is estimated to be as many as $\sim 10^{40}$ for 76-residue ubiquitin given the backbone coordinates in the PDB code $1ubq$. Allowing both side chain and backbone flexibility is likely to increase $N_{rconf}$, which probably increase exponentially with the number of residues and consequently will certainly be more intractable for larger proteins. Therefore, solving local integrals $Z_i$s within RCONFs is a not a feasible path for calculating configurational integral $Z$ of the whole conformational space, or $Z^A$ of a given macrostate $A$. A drastic simplification is to assume that all $Z_i$s for a given protein under specific solvent conditions have the same constant value $Z_{rconf}$ across the whole conformational space, and consequently:

\begin{align}
Z & = Z_{rconf}N_{rconf}(U(r_p, r_v)) \\
\Delta F^{AB} & =  k_BTln\frac{N^A_{rconf}(U(r_p, r_v)) }{N^B_{rconf}(U(r_p, r_v))} \\
S_{rconf} = k_BlnN_{rconf}(U(r_p, r_v)) \\
\Delta S^{AB}_{rconf} = k_B ln\frac{N^B_{rconf}(U(r_p, r_v)) }{N^A_{rconf}(U(r_p, r_v))}
\end{align}
Under this drastically simplified assumption of constant local configurational integral across all RCONFs, RCONFs of proteins behave like configurations of ideal gas, and change of free energy between two macrostates is equivalent to that of $S_{rconf}$ with a mere difference of a negative temperature factor. It is important to note that the hypothesized constant $Z_{rconf}$ is dependent on both identity of protein molecules and specific solvent conditions. Additionally, even under this assumption, $S_{rconf\tn{-}vib}$ for different RCONFs may vary. Since it is demonstrated that entropic contributions of hard DOFs (bonding and bending) is separable from that of soft ( torsional ) ones\cite{Li2009}, and our primary goal is to investigate how RCONFs distribute in the protein conformational space, we leave out the complexity of analyzing bonding and bending vibrational entropies in this study. 

The absolute RCONF-based conformational entropy $S_{rconf}$ of a typically-sized protein is extremely difficult, if ever possible, to be obtained by a direct sampling approach, such as MD simulations or \emph{regular biochemical experiments}, where usually micro-molar or much less proteins are used. For the above mentioned example of ubiquitin with backbone configurations fixed as the in the crystal structure $1ubq$, assuming a $10\tn{-}fs$ average life time and that each RCONF is visited only once, it will take 1 mole of ubiquitin ($\sim$ 9 kilograms) a few minutes to complete a traverse of $\sim10^{40}$ RCONFs. In reality, the measured change of conformational entropy for a typical major protein conformational change between two macrostates $A$ and $B$ is the change (presumably converged) of observed local conformational entropies.
\begin{equation}
\Delta S^{AB}_{rconf} = \Delta S^{O-AB}_{rconf}  = S^{O-A}_{rconf} - S^{O-B}_{rconf}
\end{equation} 
with $O-$ indicate observed part of the specified conformational space defining a macrostate. Again, it is important to emphasize this is not only the case for MD simulations, but also true for typical biochemical measurements and physiological activity of proteins. Since direct analysis of RCONF distributions in the whole conformational space is intractable, we take a step back and analyze $S^{O-}_{rconf}$ for arbitrarily selected region of the visited protein conformational space. The logic is that \emph{rule of RCONF distribution manifested among arbitrarily selected parts of conformational space should be effectively true for the whole conformational space}. MD trajectories provide a convenient path for arbitrary partitioning of conformational space for given protein molecules.  

It is well established in the informational theory field\cite{Shannon1948} that for a static distribution with well-defined basic states, as in the case of equilibrium distributions of RCONFs in the conformational space, entropy may be constructed by arbitrary division of the whole system into $M$ subparts. 
\begin{align}
S &= -\sum^{i=N}_{i=1} P_ilnP_i = -\sum^{j=M}_{j=1}P_jlnP_j + \sum^{j=M}_{j=1} P_jS_j \\
S_j &= -\sum^{k=k_j}_{k=1}P_klnP_k   \quad (j = 1, 2, \cdot\cdot\cdot ,M) \\
N & = \sum^{j=M}_{j=1}k_j
\end{align}
with $P_i$, $P_j$ and $P_k$ being properly normalized:
\begin{equation}
\sum^{i=N}_{i=1}P_i = 1, \quad \sum^{j=M}_{j=1}P_j = 1 \quad \tn{and} \quad \sum^{k=k_j}_{k=1}P_k = 1 \quad (j = 1, 2, \cdot\cdot\cdot ,M)
\end{equation}
$S$ is the global informational entropy and $S_j$s $(j = 1, 2, \cdot\cdot\cdot ,M)$ are local informational entropies, it is noted that such division may be carried out recursively. %By a simple multiplication of $k_B$ (Boltzmann constant), 
We may similarly divide the whole conformational space of a protein into $M$ arbitrary parts according to our need. In reality, we rarely care the global conformational entropy of a protein molecule. Instead, what we are most interested in are differences between local conformational entropies of relevant macrostates (conformations).

We carried RCONF analysis based on extensive MD trajectories of six globular proteins in native ensemble to test the ideal gas hypothesis. While this approximation does not perform well for absolute value of $S_{rconf}$ for given macrostates at very fine time resolution ($10fs$) of observation, it is demonstrated to be highly accurate and reliable, when $\Delta S_{rconf}$ or $\Delta F$ is the major concern, for all investigated protein molecules regardless of the time resolution of observation. The advantage of the end-point free energy estimation strategy as indicated by equation (23) is discussed.  

\section*{Results}
\subsection*{Ideal gas behavior of RCONFs}
To test our hypothesis, we collected extensive MD trajectories of six globular proteins with different folds and sizes. Structures of these proteins are presented in Fig. 1. By encoding torsional states of side chains into bit vectors and using the radix sorting algorithm\cite{IntroAlgorithm}, we assigned snapshots from MD trajectories to unique RCONFs, which are defined according to rotameric states as reported by Scouras and Daggett\cite{Scouras2011}, the relevant information of MD trajectories and the results are listed in Table 1. For three trajectory sets BPTI-a, CDK2 and BamC, each snapshot corresponds to a unique RCONF (i.e. $N_{rconf} = N_{snap}$). Consequently, all sampled RCONFs have the same observed probability under the given time resolutions ($250 ps$, $2 ps$ and $2 ps$) for the three proteins. Therefore, the observed RCONFs in these trajectories behave like ideal gas configurations. When the visited conformational space is divided into $M$ arbitrary partitions:
\begin{equation}
S^{O_j-}_{rconf} = -k_B\sum^{k = n^j_{rconf}}_{k=1}P_klnP_k = k_Bln(n^j_{rconf}) \quad (n^1_{rconf} + n^2_{rconf} + \cdot\cdot\cdot + n^M_{rconf} = N_{snap})
\end{equation}
with $S^{O_j-}_{rconf}$ being the observed local RCONF-based conformational entropy for the $j$th of the given $M$ arbitrary partitions. $n^j_{rconf}$ being the number of observed RCONFs in the corresponding region of conformational space. However, for other trajectory sets of different proteins (HEWL and BamE), it is apparent that $N_{rconf}$ does not equal to $N_{snap}$ anymore. To quantitatively characterize deviations from ideal gas behavior of recorded RCONFs in these trajectories, we calculated for each set of protein trajectories both the observed $S^{O-}_{rconf}$ and its ideal gas approximations $S^{O-ig}_{rconf}$ for the whole visited conformational space as shown in the equation below:
\begin{align}
S^{O-}_{rconf} &= -k_B\sum^{i = N_{rconf}}_{i=1}P_ilnP_i\\
S^{O-ig}_{rconf} &=  k_Bln(N_{rconf}) 
\end{align}
the deviation from the ideal gas approximation:
\begin{equation}
\delta S_{rconf} = S^{O-}_{rconf} - S^{O-ig}_{rconf}
\end{equation}
is listed in Table 1. Non-zero $\delta S$ were observed  for some MD trajectories with pico-second(s) ($1ps$ to $4ps$) snapshot intervals. This observation indicates that much larger deviations from ideal gas behavior may have been observed if significantly finer intervals were utilized to record MD trajectories. 

To resolve this potential concern, we generated three sets of fine resolution trajectories for proteins HEWL, BPTI and KLKA, and denoted them as HEWL-b, BPTI-c and KLKA-b respectively.  Origins of these trajectories are uniformly distributed in the corresponding set of trajectories recorded with pico-second(s) ($1ps$ to $4ps$) intervals. The interval for saving snapshots is set to 10$fs$, which is comparable with typical bonding vibrational cycles and is expected to capture interesting torsional transitions within the visited conformational space. At this time scale resolution, $N_{rconf}$ is indeed quite different from $N_{snap}$ and significantly larger $\delta S$s are observed (Table 1.). We further examined the statistical weight ($w_{rconf}$) of RCONFs, which under the assumption of the constant local integral $Z_{rconf}$ should be a constant across all RCONFs. As shown in Fig. ~\ref{w-rconf}, probability of RCONFs for HEWL follows approximately an exponential decay as a function of $w_{rconf}$. These observations suggest that the ideal gas hypothesis might not be helpful in dealing with protein conformational distributions.  

However, what we care the most is the change of observed local conformational entropy ($\Delta S^{O-}_{rconf}$) and free energy between macrostates in cases of interested events (e.g. conformational change or molecular binding). To analyze behavior of $\Delta S^{O-}_{rconf}$ between arbitrary partitions of conformational space visited by $10\tn{-}fs$ interval trajectories of HEWL, we chose the following different ways of conformational space division. Firstly, we take a given backbone dihedral ($\phi$ or $\psi$) as the order parameter and divide the whole visited conformational space into 20 windows on it. $S^{O_j-}_{rconf}$ and $S^{O_j-ig}_{rconf} (j = 1,2,\cdot\cdot\cdot,20)$ were calculated for each window. For each backbone torsion, such division and calculation was performed, and a $S^{O_j-}_{rconf}$ vs. $S^{O_j-ig}_{rconf}$ plot was generated. Strong linear correlations were observed for all 256 plots. After performing linear fit, the distributions of slopes and correlation coefficients are shown in Fig. ~\ref{slope}. Essentially, both the slope and correlation coefficient are approximately equal to 1, indicating the robustness of the ideal gas behavior as far as the change of $S_{rconf}$ is concerned.

The above mentioned plots are constructed for sets of non-overlapping and complete conformational partitions. A set of $n$ partitions ($\Omega_i, i = 1, 2, \cdot\cdot\cdot, n$) in a specified configurational space $\Omega$ are non-overlapping and complete if:
\begin{align}
\Omega_i \cap \Omega_j & = \emptyset \quad(i \ne j \quad \tn{and} \quad i, j = 1, 2, \cdot\cdot\cdot, n) \\
\tn{and}\quad \Omega & = \Omega_1 \cup \Omega_2 \cup \cdot\cdot\cdot \cup \Omega_n
\end{align}
Since what we want to analyze are distributions of RCONFs in arbitrary conformational partitions, which certainly may overlap. In reality, it is not unusual for two interested macrostates to overlap in conformational space (e.g. ligand bound and ligand free proteins for a given protein-ligand system), and such overlap is the theoretical foundation of the well-acknowledged conformational selection mechanism\cite{CSIF}. We therefore constructed a $S^{O_j-}_{rconf}$ vs. $S^{O_j-ig}_{rconf}$ plot for all partitions based on various backbone torsions, as shown in Fig. 4. %~\ref{sovssid}. 
Again, both the slope and the correlation coefficient of a linear fit are approximately being 1.0.

Next we considered two alternative order parameters for conformational space partition that may not be constructed as linear combinations of backbone torsional DOFs, radius of gyration $R_g$ and number of native contacts $N_{nc}$ (see ref\cite{hewl} for specific definition of native contacts of HEWL). After these two quantities were calculated for each snapshot in the $10\tn{-}fs$ interval HEWL trajectories, $20$ equal-width windows were created for both $R_g$ and $N_{nc}$, $S^{O_j-}_{rconf}$ vs. $S^{O_j-ig}_{rconf}$ plot was constructed for the $40$ conformational partitions as shown in Fig. 4. %~\ref{sovssig}. 
Similarly, both the slope and the correlation coefficient of a linear fit are approximately being 1.0. The observation that the slope being approximately  1.0 in these plots implies that $\Delta S^{O-} \approx \Delta S^{O-ig}$. Therefore, although ideal gas behavior of $S_{rconf}$ is only true for sufficiently coarse time resolution of observation, it is a very good approximation for $\Delta S_{rconf}$ regardless of time resolution of observation. 

\subsection*{Equivalence between RCONF based conformational entropy and free energy}

As shown in equations 23 and 25, equivalence between $\Delta S^{AB}_{rconf}$ and $\Delta F^{AB}$ (except the negative temperature factor) is another major conclusion of the ideal gas hypothesis. In an equilibrium canonical system, free energy difference between two given part of conformational space may be effectively calculated based upon observed populations.    
\begin{equation}
\Delta F^{AB} = k_BTln \frac{N^A}{N^B} 
\end{equation}
with $A$ and $B$ stands for two arbitrarily given partitions of conformational space (macrostates), and $N^A$ and $N^B$ being observed populations, which are effectively represented by the number of snapshots ($N^A_{snap}$ and $N^B_{snap}$) in equilibrium MD trajectories. Imagining that there is a reference conformation in equilibrium with other visited parts of conformational space, and this conformation has a statistical weight corresponding to $1$ snapshot, then relative free energy of any given part of conformational space $A$ becomes $-k_BTln N^A_{snap}$, with $N^A_{snap}$ being the number of snapshots in $A$.

We plotted relative free energy $-k_BTlnN_{snap}$ as a function of local ideal gas entropy $S^{O-ig}_{rconf} = k_Bln N_{rconf}$ for the above mentioned sets of conformational space partitions (see Fig. 5.)% ~\ref{fevssid}). 
 A strong linear correlation is observed with a slope of approximately -$1.0$ regardless of different ways of conformational space partitioning. These observations further validated the ideal gas hypotheses. To utilize equation (23) as a new end-point free energy calculation method, there are two possible paths. The first one is to directly calculate the ratio $\frac{N^A_{rconf}(U(r_p, r_v))}{N^B_{rconf}(U(r_p, r_v))}$ without knowing the absolute values of both the numerator and the denominator. This is convenient when converged equilibrium sampling of conformations $A$ and $B$ is readily achievable. Direct calculation of both $N^A_{rconf}(U(r_p, r_v))$ and $N^B_{rconf}(U(r_p, r_v))$ provides a potential alternative when converged sampling of both end states are difficult. However, as briefly mentioned in the introduction, due to astronomically large number of RCONFs for typically sized proteins, enumeration of RCONFs is not realistic for a specified region of conformational space from regular MD simulations, from enhanced sampling techniques, or from typical biochemical experiments. On the other hand, importance sampling\cite{zhang2006} in combination with sequential Monte Carlo\cite{zhang2003} was demonstrated to be an efficient and reliable way of counting the number of conformers for fixed backbone and rigid side chain rotamers.  With incorporation of side chain and backbone flexibility, the number of RCONFs may be effectively counted by such importance sampling procedures. The most appealing feature of this approach is that no overlapping of conformational space is required. Additionally, such calculation will be compatible with solvation treatment ranging from full explicit solvent to simple implicit solvation models and anything in between. 

Similar validation analyses of ideal gas hypothesis are performed for all trajectories with a non-zero $\delta S_{rconf}$ and the same conclusion is reached (data not shown). 
%Results for two sets of BPTI trajectories (BPTI-b, $\Delta t = 1 ps$; BPTI-b, $\Delta t = 0.01 ps$) were provided in the supporting info (Fig. S1 and Fig. S2).

\subsection*{Analysis with alternatively defined basic conformers}
Most trajectories utilized in this study are generated with CHARMM22 force fields, except for the fact that CDK2 trajectories are generated with AMBER ff12SB and BPTI-a trajectories with a modified AMBER ff99SB\cite{Shaw2010}. Differences in distributions of side chain torsional DOFs are expected between MD trajectories generated by different force fields, and between experimental structures and MD trajectories of given force fields. Based on distributions of heavy-atom-defined side chain torsional angles observed in each set of trajectories, we utilized an in-house torsional state assignment procedure\cite{hewl}  to define trajectory-set specific basic conformers, termed RCONF2s. We compared the results from HEWL trajectory set with what reported by Scouras and Daggett\cite{Scouras2011}, which was used to define RCONFs. While most boundaries for torsional state assignment only differ for a few degrees, a number of torsional DOFs in five residues (GLU, GLN, ASN, ASP and ARG) were given significantly different definition of torsional states as shown in Fig. S3. The point here is not to raise an argument regarding the optimal way of assigning torsional states for side chain torsions. Rather, the differences give us an opportunity to test the sensitivity of the ideal gas hypothesis on the specific definition of basic conformers. We repeated the analysis for RCONF2s. Corresponding $N_{rconf2}$ and $\delta S_{rconf2}$, which are different from $N_{rconf}$ and $\delta S_{rconf}$ for some trajectory sets, were shown in Table 1. Nonetheless, the relationship between $S^{O-}_{rconf2}$ and $S^{O-ig}_{rconf2}$, and that between $F$ and $S^{O-ig}_{rconf2}$ is essentially the same as what observed for RCONFs, the results are shown only for HEWL-b trajectory set (See Fig. 6.).%~\ref{rconf2}). 
%Additional analysis of trajectories based on definition of basic conformers according to experimental rotameric libraries\cite{Shapovalov2011} resulted in the same conclusion (data not shown). 
  These observations indicate that the ideal gas hypothesis is not sensitive to details in the specific definition of basic conformers.     
 
\section*{Discussions}
We did not include backbone torsional DOFs in the definition of RCONF(2)s based on the following consideration. Firstly, in protein folding, design and docking studies, backbone DOFs are usually treated explicitly, and free energy difference between two given backbone configurations are estimated with various scoring functions. Consistent with the idea that folded proteins have solid like backbone and liquid like side chains\cite{Kresten2005}, it is found in our previous backbone conformational analysis of HEWL that the number of statistically significant combinations of backbone torsional states is very limited\cite{hewl}. How to pick right backbone configurations out of astronomically large possible number of which for explicit free energy estimation is another difficult task to tackle in predictive tasks such as folding, design and docking, and we are actively investigating this issue. In our conformational partitions based on individual backbone dihedrals, backbone DOFs are not limited except for the one on which the projection is performed. No explicit restriction of backbone torsional DOFs is imposed on conformational partitions based on radius of gyration $R_g$ and on number of native contacts $N_{nc}$. The observed validity of the ideal gas hypothesis indicates that change of backbone torsional states are effectively reflected by change of relevant side chain torsional states, at least in a statistical sense. 

$S_{rconf\tn{-}vib}$ in equation 14 includes contributions from both bonding and bending vibrational entropies, and from local part of torsional entropy within each RCONF. Seemingly, each RCONF allow significant local torsional motion since each defining torsion angle have a $\sim120$ (or $\sim60$ in a few cases) degree range to fluctuate.  However, analysis of high resolution trajectories in sets HEWL-b and BPTI-c indicate this is not the case, the average life time of RCONFs are $\sim 2.3fs$ and $\sim 2.1fs$ respectively. This is due to the fact that even a small local torsional motion in one side chain torsional DOF is likely to be accompanied by change of torsional state in some other side chain torsions, the large number of side chain torsional DOFs results in short life time of RCONFs and correspondingly highly limited local torsional motion in each RCONF. In this study, we focused exclusively on conformational entropy based on RCONFs. The corresponding vibrational contributions to the configurational entropy ( $S_{rconf\tn{-}vib}$ ), together with other essential components of free energy in equations (2-3), were not analyzed. It is certainly desirable to have the capability to nail down these terms with high level of confidence. Unfortunately, reliable calculation of $S_{rconf\tn{-}vib}$ is difficult by quasiharmonic or correlation based methods for most conformers due to limited number of snapshots available. No effective methods is presently available for calculation of correlation entropies between different type of molecular components ($S_{pl}$, $S_{pv}$ and $S_{lv}$) in protein-ligand and other similar type of systems. 

At first sight, the effective equivalence between change of conformational entropy and change of free energy seems exotic, and one would wondering what happened to enthalpic contributions and vibrational entropic contributions. It is important to note that the seemingly only important quantity $N_{rconf}$ is a function of underlying molecular interactions including both intramolecular interactions within a protein molecule and intermolecular interactions between protein and solvent. Additionally, we emphasize that the constant local integral assumption does not necessarily limit $S_{rconf\tn{-}vib}$, which might vary significantly for different RCONFs. We speculate that the silence of $S_{rconf\tn{-}vib}$ in the observed change of free energy might due to its correlation, and consequently canceling effects, with other complex terms in equations 2 and 3.

The proposed end point free energy estimation methodology, as shown in equation 23, is in principle complementary to presently widely utilized methods such as Linear Interaction Energy (LIE) model\cite{LIE1998} and MM/P(G)BSA \cite{MMPBSA-JMC}, especially for the cases where virtually no overlapping of conformational space exist for two end macrostates. Another advantage of equation 23 is that there is no system dependent parameters to construct. It was demonstrated that side chain conformational entropies for given backbone configurations are not sensitive to force fields details\cite{zhang2006}. However, the reported results are restricted to backbone of folded proteins or decoys that have reasonable packing density and surface exposure. To use equation 23 alone for selecting proper backbone configurations, the quality of solvation model is likely to be of critical importance.

United atom model is widely utilized to improve computational efficiency\cite{Berger1997,Chiu2009,Tj2014}. Since RCONFs (or RCONF2s) are defined by heavy atoms, properly parameterized united atom models may potentially be utilized for counting number of RCONFs without significantly compromising accuracy. The present study is limited to native globular proteins. The physiological importance of membrane proteins and inherently disorder proteins are well acknowledged, some proteins interconvert between folded and unfolded states many times during their physiological life time. We are working on the generalization of the ideal gas hypothesis to these widely different scenarios, and to other complex molecular systems as well.

\section*{Conclusion}
In summary, we proposed the ideal gas hypothesis to deal with lack of fundamental mircostates in defining classical entropy of proteins. By utilizing the expediency of extensive MD trajectories in analyzing arbitrary partitions of protein conformational space, we tested the ideal gas hypothesis of RCONFs for a few globular proteins in native ensemble. The ideal gas hypothesis, while performs poorly for estimating absolute value of conformational entropy when the time resolution of observation is sufficiently fine, is demonstrated to be consistently robust as far as change of conformational entropy (or free energy) is concerned. A new end point free energy estimation method, which is a direct result of the ideal gas hypothesis, is also examined. This alternative free energy estimation scheme is applicable to cases where end states do not overlap in conformational space, which are highly challenging situations for presently available free energy methodologies.

\section*{Methods}
BPTI-a trajectories were provided by DE Shaw\cite{Shaw2010}. MD trajectories of HEWL ( collectively $200\mu s$ comprising 2000 $100\tn{-}ns$ trajectories) were taken from our previous simulation study\cite{hewl}. BPTI-b and KLKA trajectories (with pico-second resolution) were taken from another previous study\cite{Wenzhao149}. BamC, BamE and CDK2 trajectories were generated in our group and details of these trajectories will be published in the future. For BamC, structure with PDB code 3TGO was solvated in 13736 water molecules, 39 $Cl^-$ and 43 $N_a$ ions. For BamE, structure with PDB code 2YH9 was solvated with 7391 water molecules, 22 $Cl^-$ and 21$N_a$ ions. CHARMM22 force fields are used for simulations of BamC and BamE. CDK2 trajectories are based on AMBER ff12 force fields, 66 crystal structures (1FIN, 1GZ8, 1HCK, 1JST, 1JST, 1PF8, 1PW2, 1PXI, 1PXJ, 1PXK, 1PXM, 1W8C,1Y8Y, 2A4L, 2B54, 2BPM, 2BPM, 2C4G, 2C4G, 2C5N, 2C5N, 2C5O, 2C5O, 2C5V, 2C5X, 2C69, 2C6K, 2C6L, 2CLX, 2EXM, 2UUE, 2V22, 2VTL, 2VTM, 2VTR, 2WEV, 2WFY, 2WHB, 2WIH, 2WIH, 2WPA, 2WXV, 2WXV, 2X1N, 2X1N, 3F5X, 3F5X, 3IGG, 3LFQ, 3PXF, 3PXZ, 3QHR, 3QQF, 3QQJ, 3RK9, 3RKB, 3S0O, 3UNK, 3WBL, 4BCK, 4EZ7, 4GCJ, 4I3Z, 4II5, 4KD1) are utilized to start 66 trajectories after each was solvated with 13851 water molecules, 49 $Cl^-$ and 45 $N_a$ ions, and production runs of $\sim 200\tn{-}ns$ are performed for each CDK2 system after equilibration. The same equilibration procedures as used in a previous study\cite{Wenzhao149} was utilized for equilibration of these three protein simulation systems. The starting structural state of all $10\tn{-}fs$ resolution trajectories are uniformly picked from corresponding pico-second(s) resolution trajectories. Specifically, 21101, 6124 and 4204 $10\tn{-}ps$ trajectories are generated for sets HEWL-b, BPTI-c and KLKA-b. 

\begin{acknowledgement}
This research was supported by National Natural Science Foundation of China under grant number 31270758. Computational resources were partially supported by High Performance Computing Center of Jilin University, China. We thank DE Shaw Research for providing BPTI trajectories. We thank Zhonghan Hu for critical reading of the manuscript.
\end{acknowledgement}

%\begin{suppinfo}
%Three figures were provided in the supporting info.
%\end{suppinfo}

\bibliography{Total}

\newpage
\begin{table}
\begin{tabular}{|c|c|c|c|c|c|c|c|c|}
\hline
Protein & $N_{res}$ & $\Delta t $ & $N_{snap}$ & $N_{tor}$ & $N_{rconf}$&$\delta S_{rconf}$ &$N_{rconf2}$& $\delta S_{rconf2}$ \\
\hline\\
BPTI-a&58&250&4120838&113&4120838&0.00 &4120815&0.000007\\
\hline\\
HEWL&129&4&50000000&196&49990573&0.000262&49668320&0.009333\\
\hline\\
CDK2&298&2&6560590&565&6560590&0.00& 6560590&0.00\\
\hline\\
BamC&190&2&943505&339&943505&0.00&921630&0.033388\\
\hline\\
BamE&68&2&1679968&120&1674297&0.004708&1094439&0.639567\\
\hline\\
KLKA&223&1&3741963&413&3741963&0.009684&3653756&0.034387\\
\hline\\
BPTI-b&58&1&3560127&113&3523927&0.014257&1425354&1.330897\\
\hline\\
KLKA-b&223&0.01&4204000&413&4201317&0.000884&823555&2.326272\\
\hline\\
HEWL-b&129&0.01&21101000&196&9194924&1.214137&2845936&2.837891\\
\hline\\
BPTI-c&58&0.01&6124000&113&2923546&1.074068&76140&5.477714\\
\hline
\end{tabular}
\caption{The list of the studied protein MD trajectory sets. For the trajectory sets of the same protein with different resolutions, we use $\tn{-}a$, $\tn{-}b$ and $\tn{-}c$ to make the distinction. $N_{res}$: number of residues; $N_{tor}$: number of heavy-atom side chain torsional DOFs utilized in defining RCONF(2)s; $\Delta t$: time interval for saving MD snapshots in $ps$; $N_{snap}$: total number of snapshots in the given trajectory set; $N_{rconf}$: total number of RCONFs in the given trajectory set; $N_{rconf2}$: total number of RCONF2s in the given trajectory set; $\delta S_{rconf}$ (in the unit of $k_B$): deviation of RCONF entropy from the ideal gas value as indicated in equation 34; $\delta S_{rconf2}$ (in the unit of $k_B$): deviation of RCONF2 entropy from the ideal gas value, defined similarly with $\delta S_{rconf}$.}
\end{table}

\newpage
\begin{figure}[] 
\centering 
\subfloat[KLKA]{\includegraphics[width=1.5in]{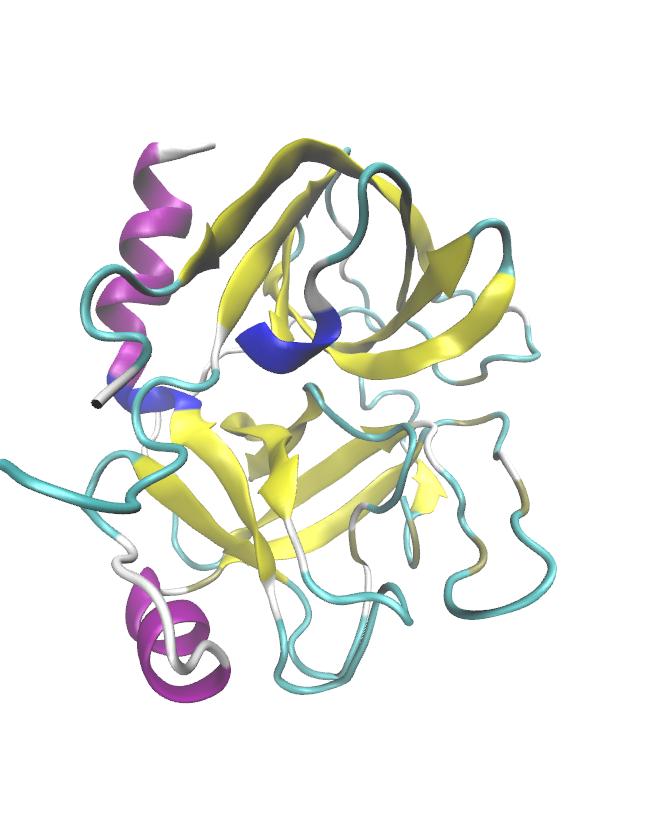}}
%\subfloat[BPTI-a]{\includegraphics[width=1.5in]{./BPTI_A_1.jpg}}\quad
\subfloat[BPTI]{\includegraphics[width=1.5in]{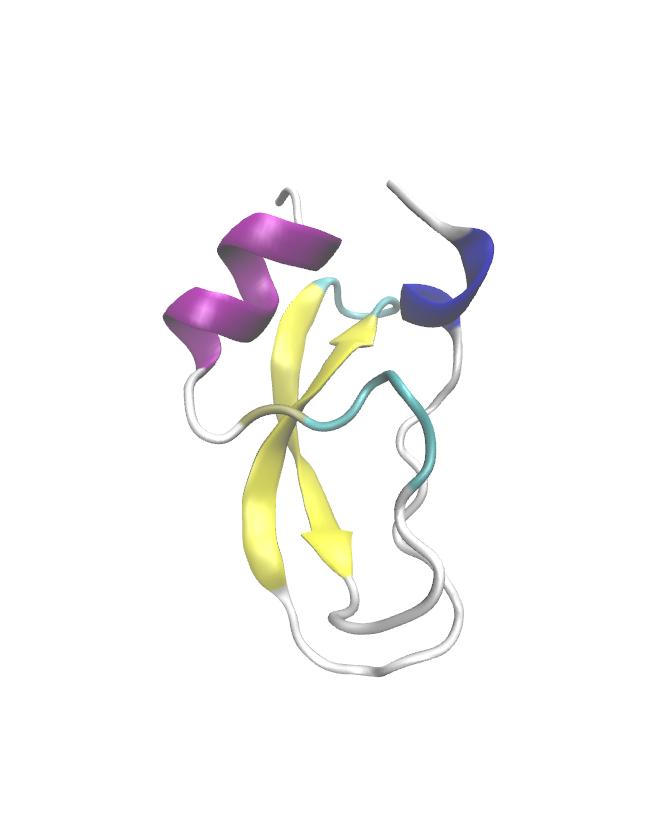}}
\subfloat[CDK2]{\includegraphics[width=1.5in]{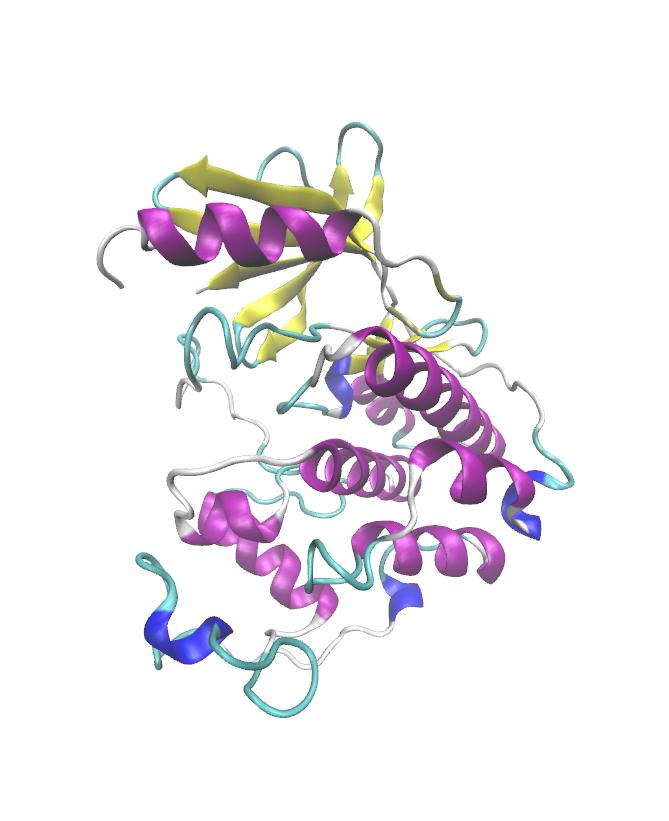}}\\
\subfloat[BamC]{\includegraphics[width=1.5in]{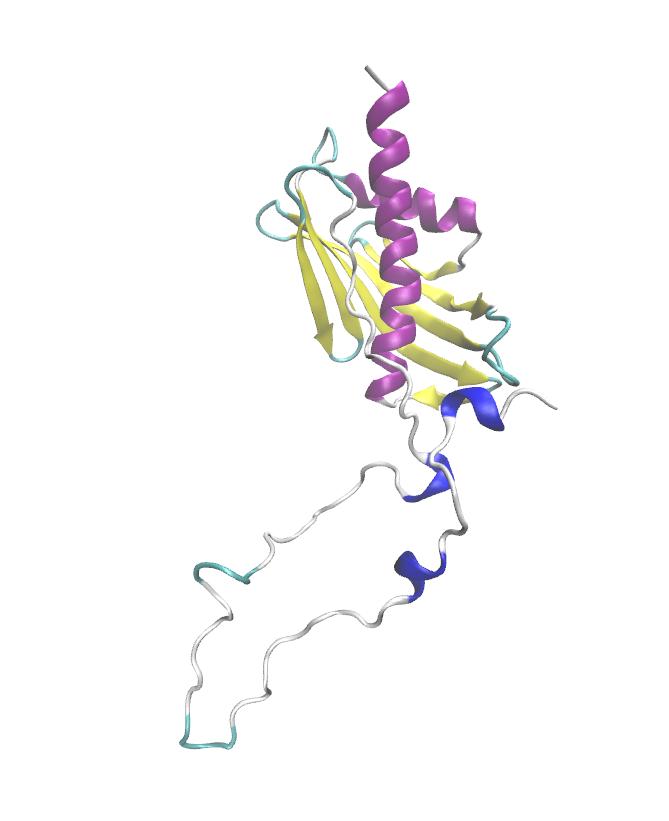}}
\subfloat[BamE]{\includegraphics[width=1.5in]{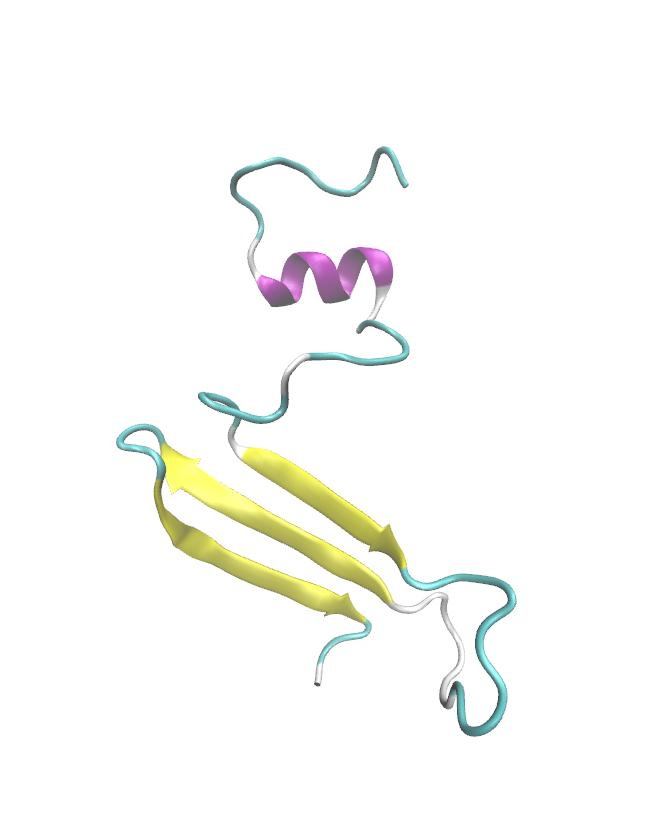}}
\subfloat[HEWL]{\includegraphics[width=1.5in]{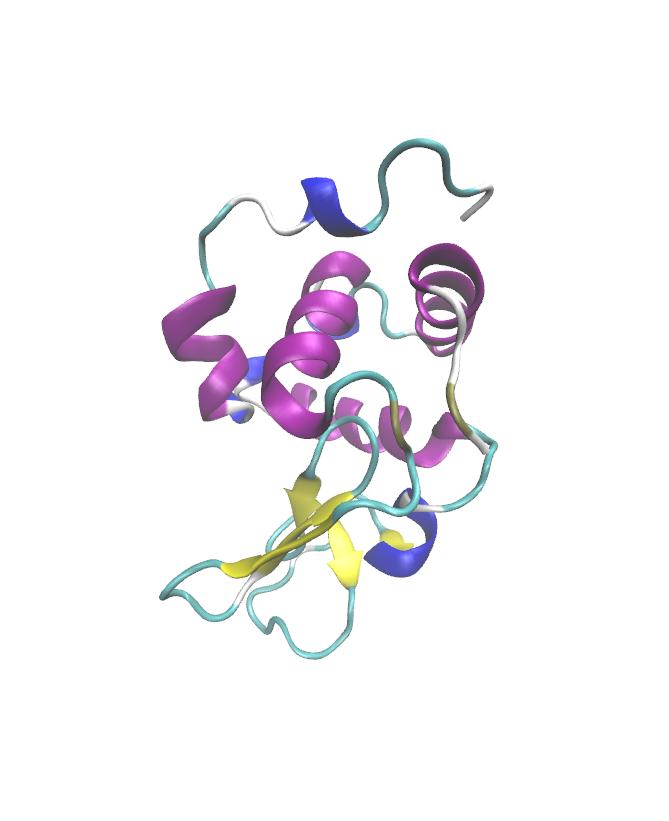}}
\caption{Structures of proteins utilized in this study. $\alpha$ helices are in purple, $\beta$ strands are in yellow, $3\tn{-}10$ helices are in blue, loops are in cyan and turns are in white. KLKA (Porcine Pancreatic Kallikrein A);  BPTI (Bovine Pancreatic Trypsin Inhibitor); CDK2 (Cyclin-dependent kinase 2); Bam (The $\beta$-barrel assembly machine); HEWL (Hen Egg White Lysozyme). Graphics are prepared using VMD.} 
\label{fig:protgraph}
\end{figure}

\newpage
\begin{figure}
\centering
\subfloat[]{\includegraphics[width=3.in]{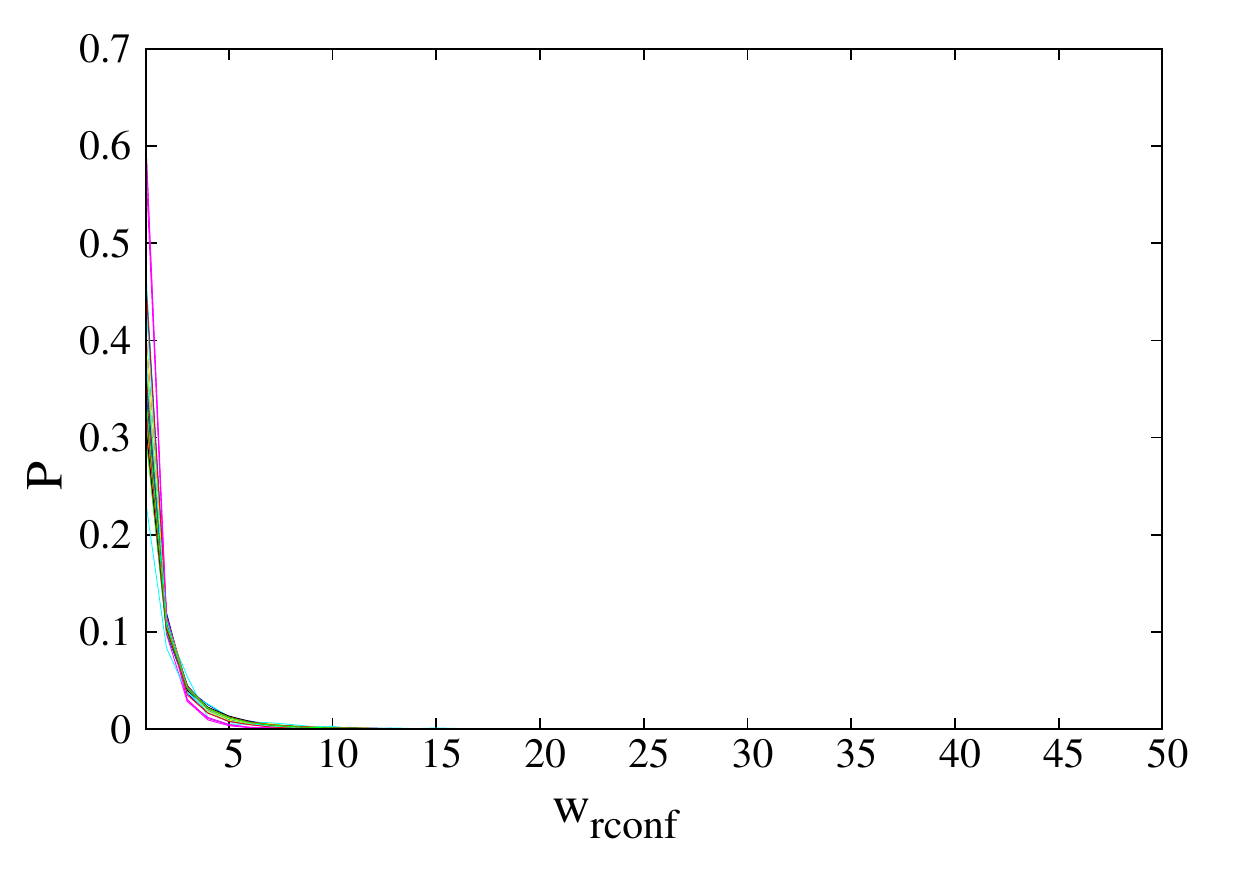}}
\subfloat[]{\includegraphics[width=3.in]{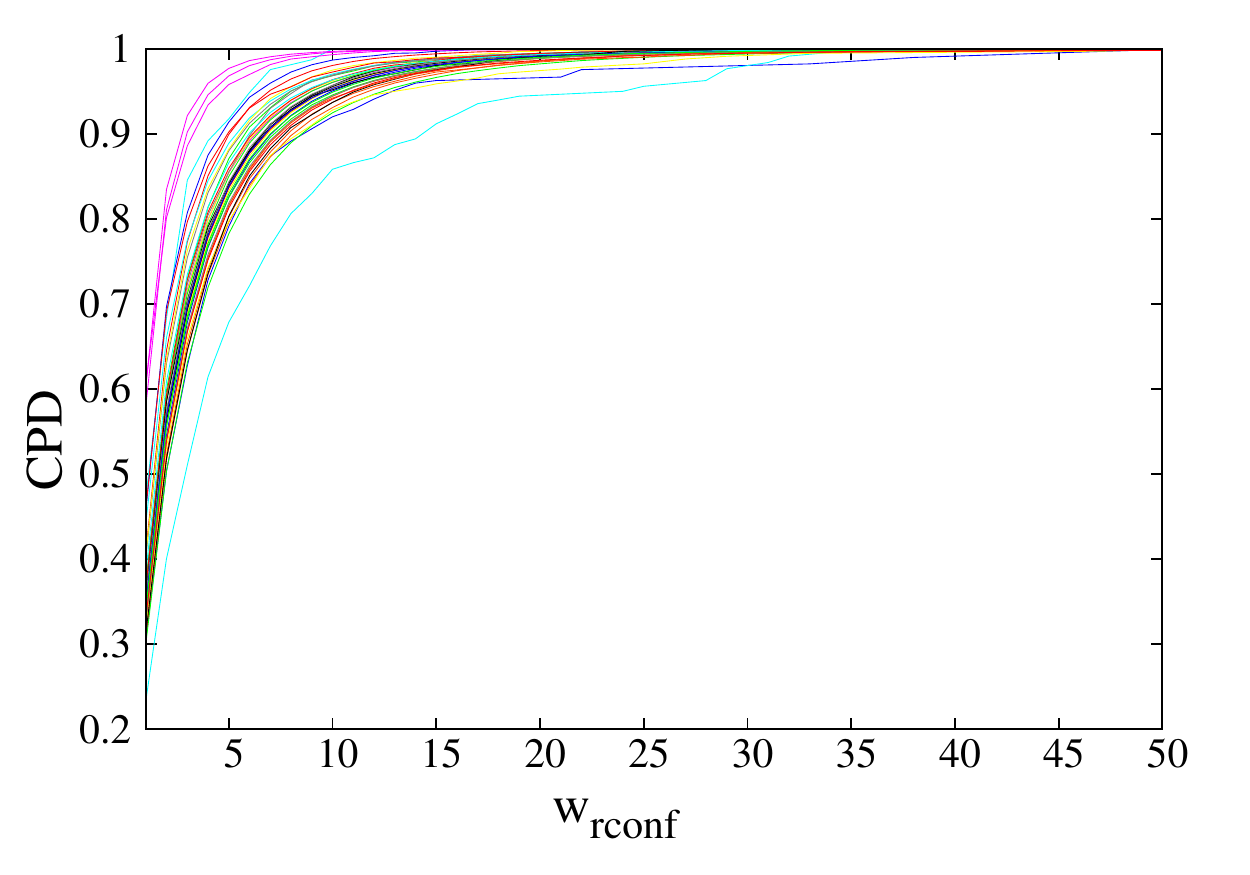}}
\caption{Statistical weight of RCONFs ($w_{rconf}$) as represented by the number of snapshots in $10\tn{-}fs$ interval HEWL trajectories. a) Probability ($P$) of various $w_{rconf}$ in 37 representative conformational partitions. b) Cumulative probability density ($CPD$) of $w_{rconf}$ for the same set of conformational partitions as in a).}
\label{w-rconf}
\end{figure}

\newpage
\begin{figure}
\centering
\subfloat[]{\includegraphics[width=3.in]{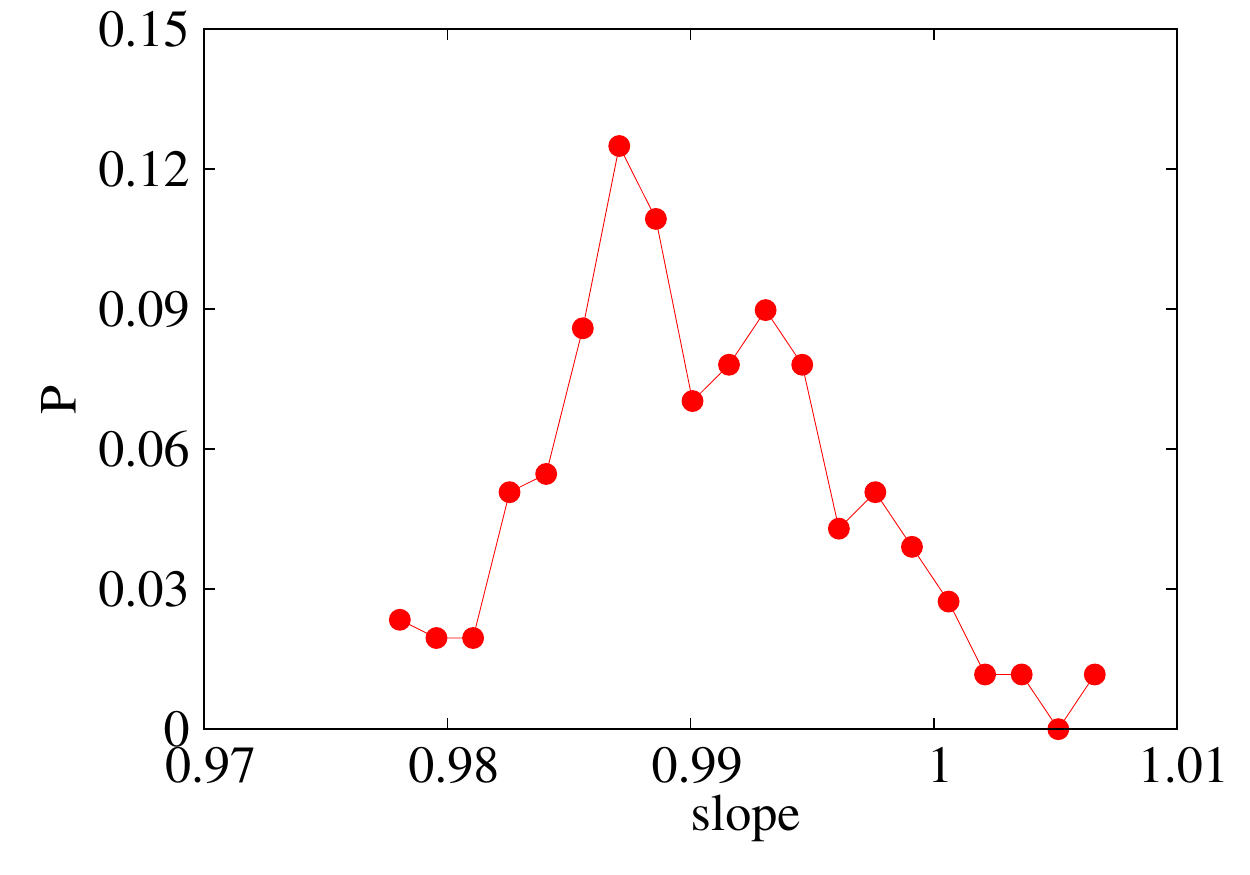}}
\subfloat[]{\includegraphics[width=3.in]{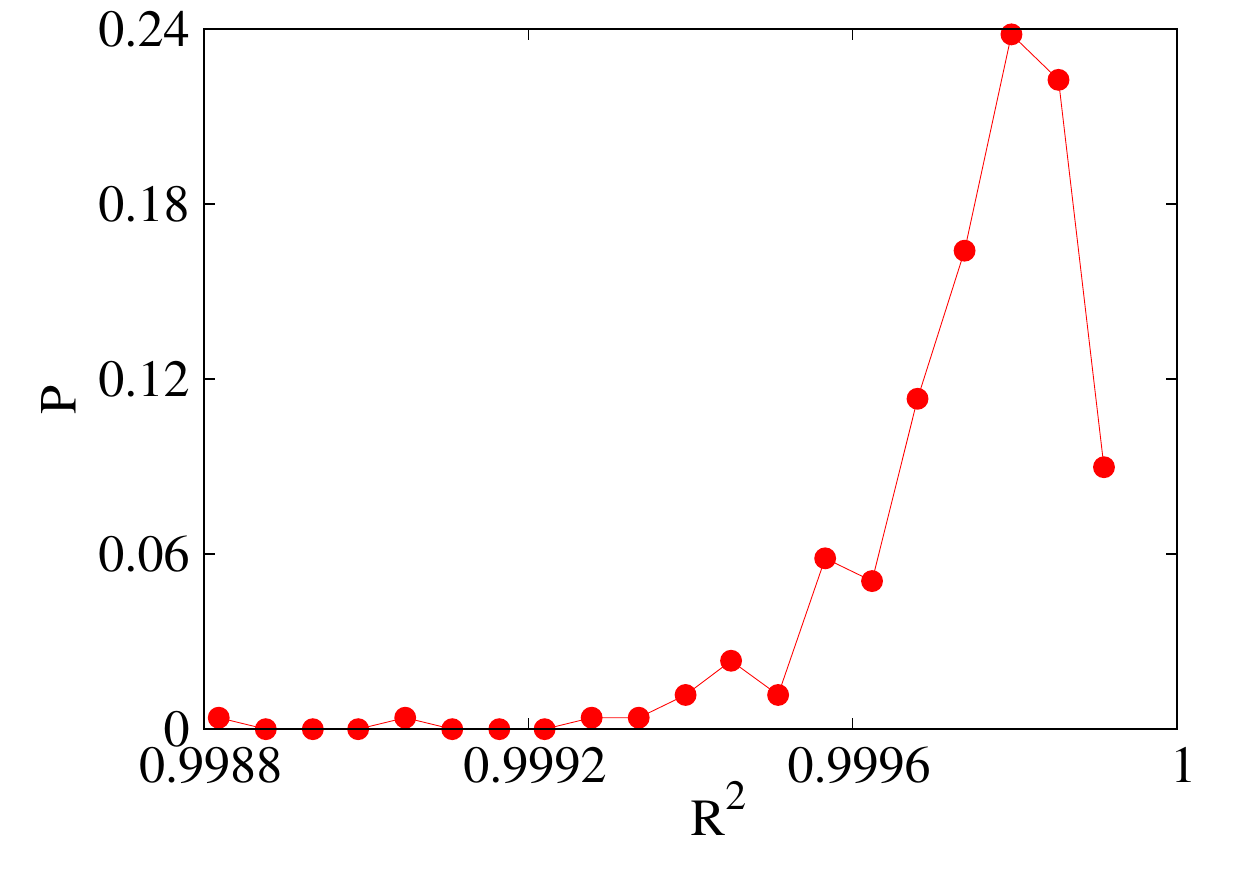}}
\caption{Probability ($P$) distributions of a)slopes and b)squared linear correlation coefficients calculated from 256 $S^{O-}_{rconf}$ vs. $S^{O-ig}_{rconf}$ plots based on 256 different ways of conformational partitions performed on HEWL-b trajectory set. Each way of conformational partition corresponds to projection of all snapshots onto a given backbone dihedral ($\phi$ or $\psi$).}
\label{slope}
\end{figure}

\newpage
\begin{figure}
\centering
\subfloat[]{\includegraphics[width=3.in]{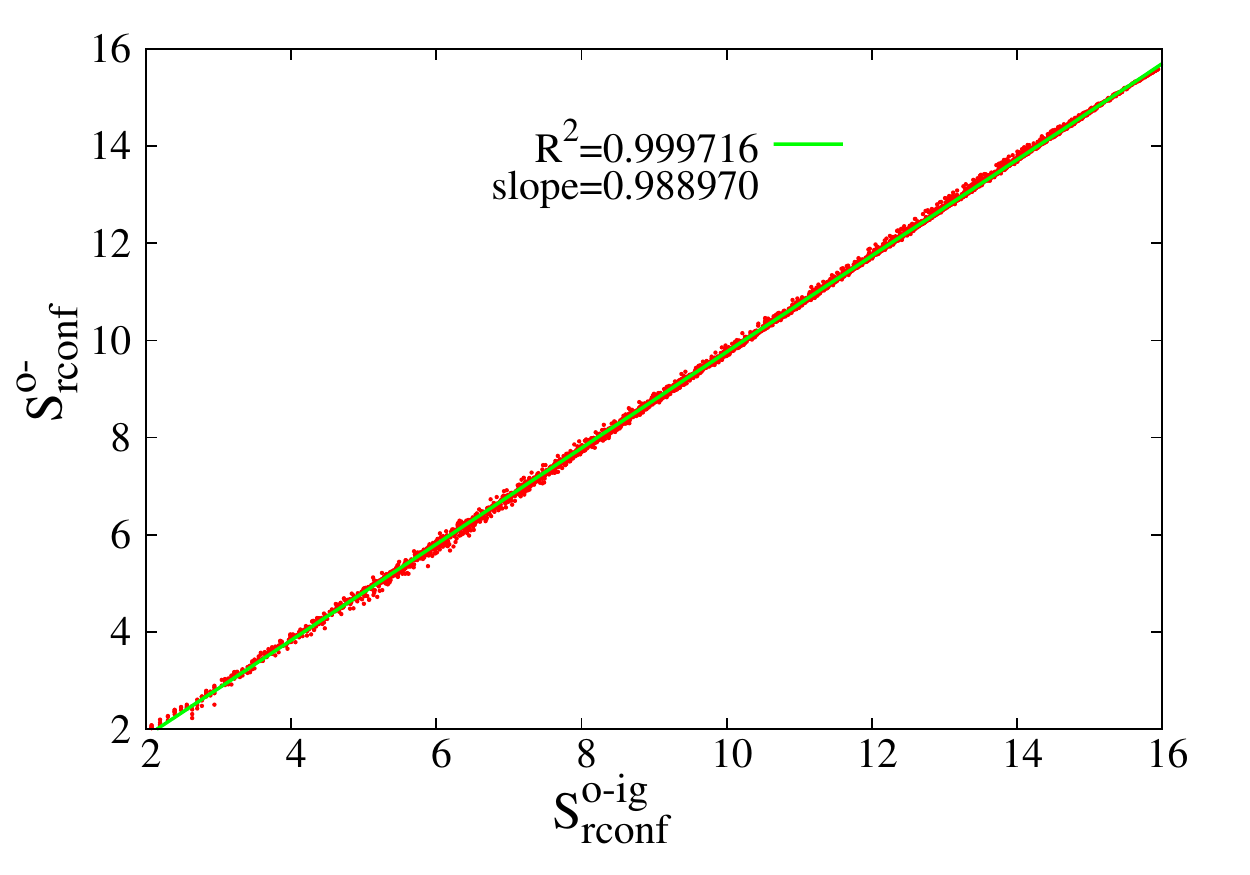}}
\subfloat[]{\includegraphics[width=3.in]{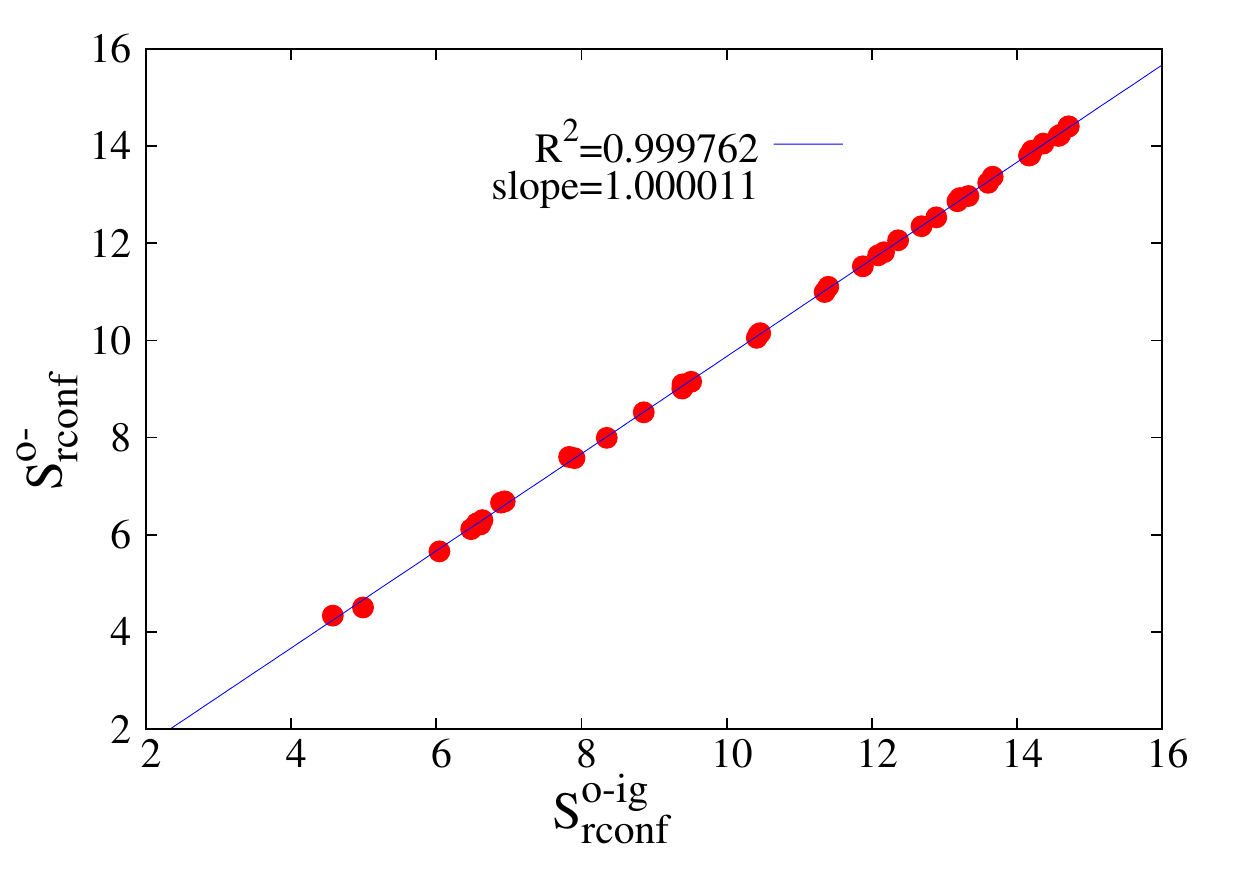}}
\caption{$S^{O-}_{rconf}$ vs. $S^{O-ig}_{rconf}$ (both in the unit of $k_B$) plots of HEWL-b for two different types of overlapping conformational partitions. a)For 5120 partitions generated from projection onto 256 backbone dihedrals. b)For 40 partitions generated from projection onto radius of gyration $R_g$ and the number of native contacts $N_{nc}$.}
%\label{sovssid}
\end{figure}

\newpage
\begin{figure}
\centering
\subfloat[]{\includegraphics[width=3.in]{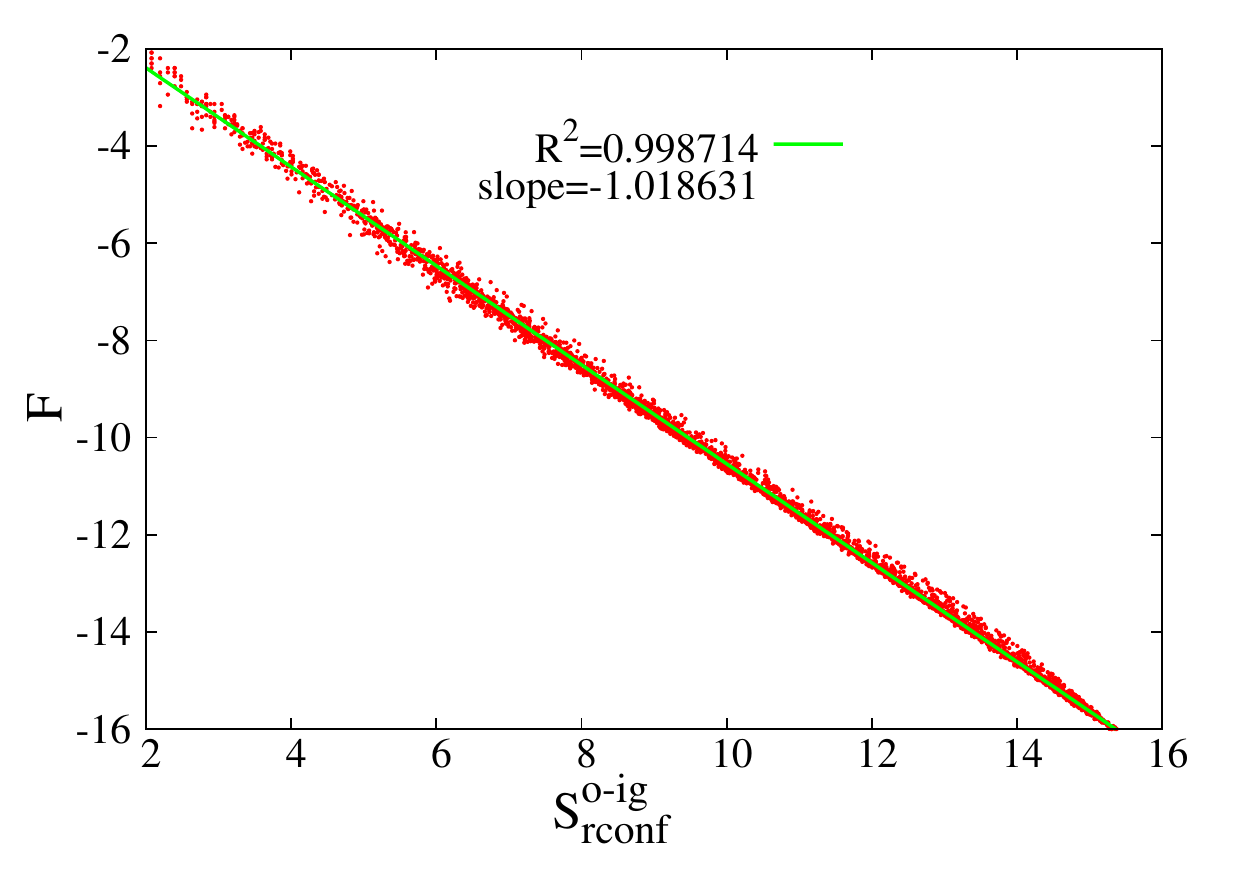}}
\subfloat[]{\includegraphics[width=3.in]{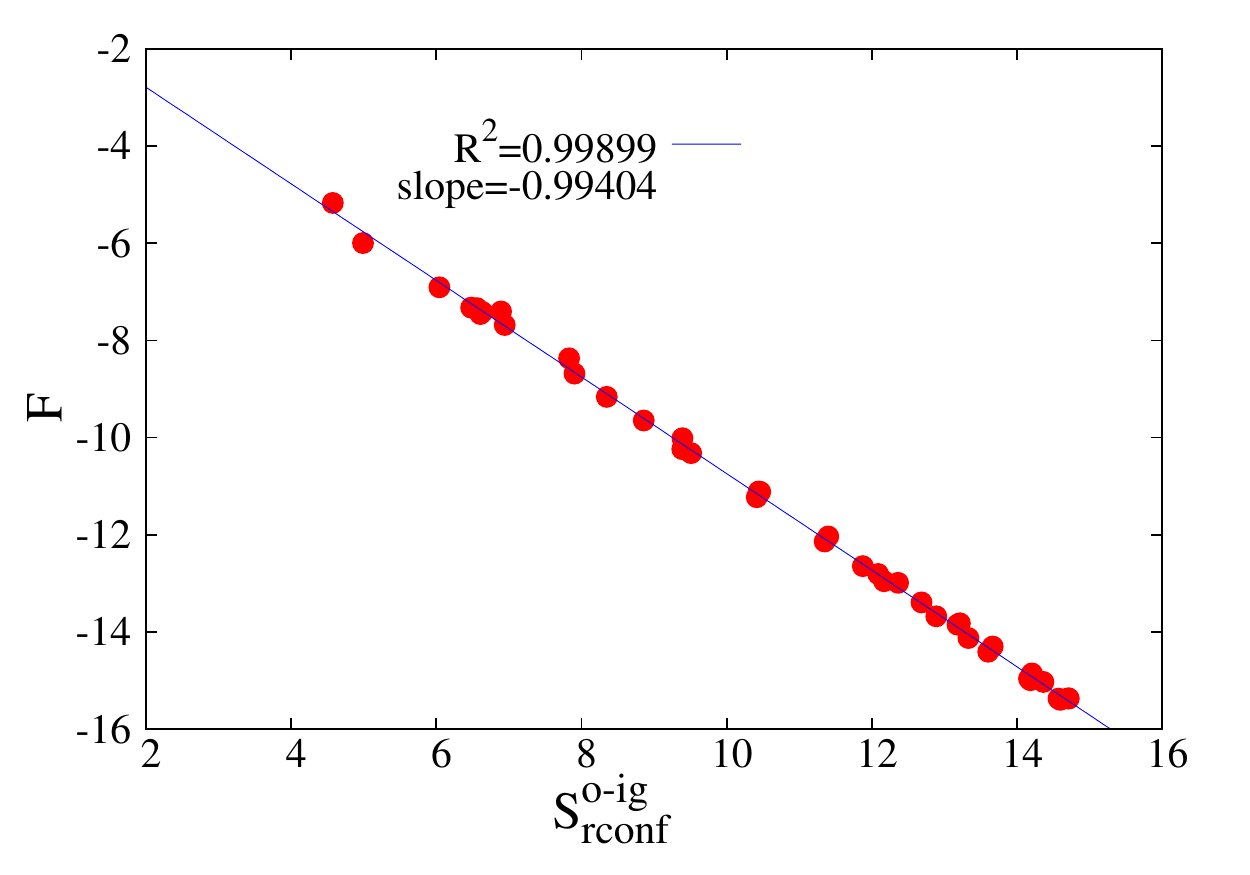}}
\caption{Relative free energy $F$ (in the unit of $k_BT$) vs. $S^{O-ig}_{rconf}$ (in the unit of $k_B$) plots for two different types of overlapping conformational partitions. a)For 5120 partitions generated from projection onto 256 backbone dihedrals. b)For 40 partitions generated from projection onto radius of gyration $R_g$ and the number of native contacts $N_{nc}$.}
%\label{fevssid}
\end{figure}

\newpage
\begin{figure}
\centering
\subfloat[]{\includegraphics[width=3.in]{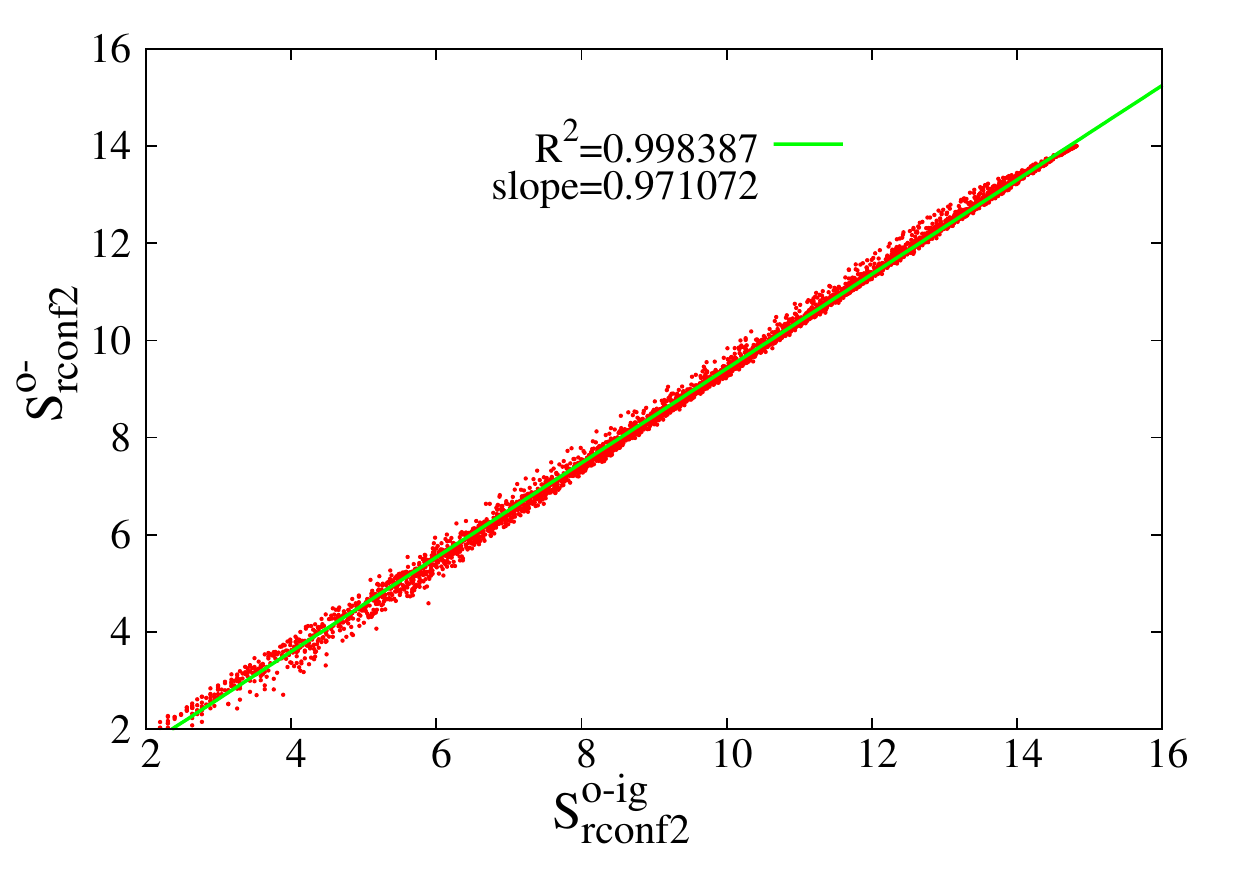}}
\subfloat[]{\includegraphics[width=3.in]{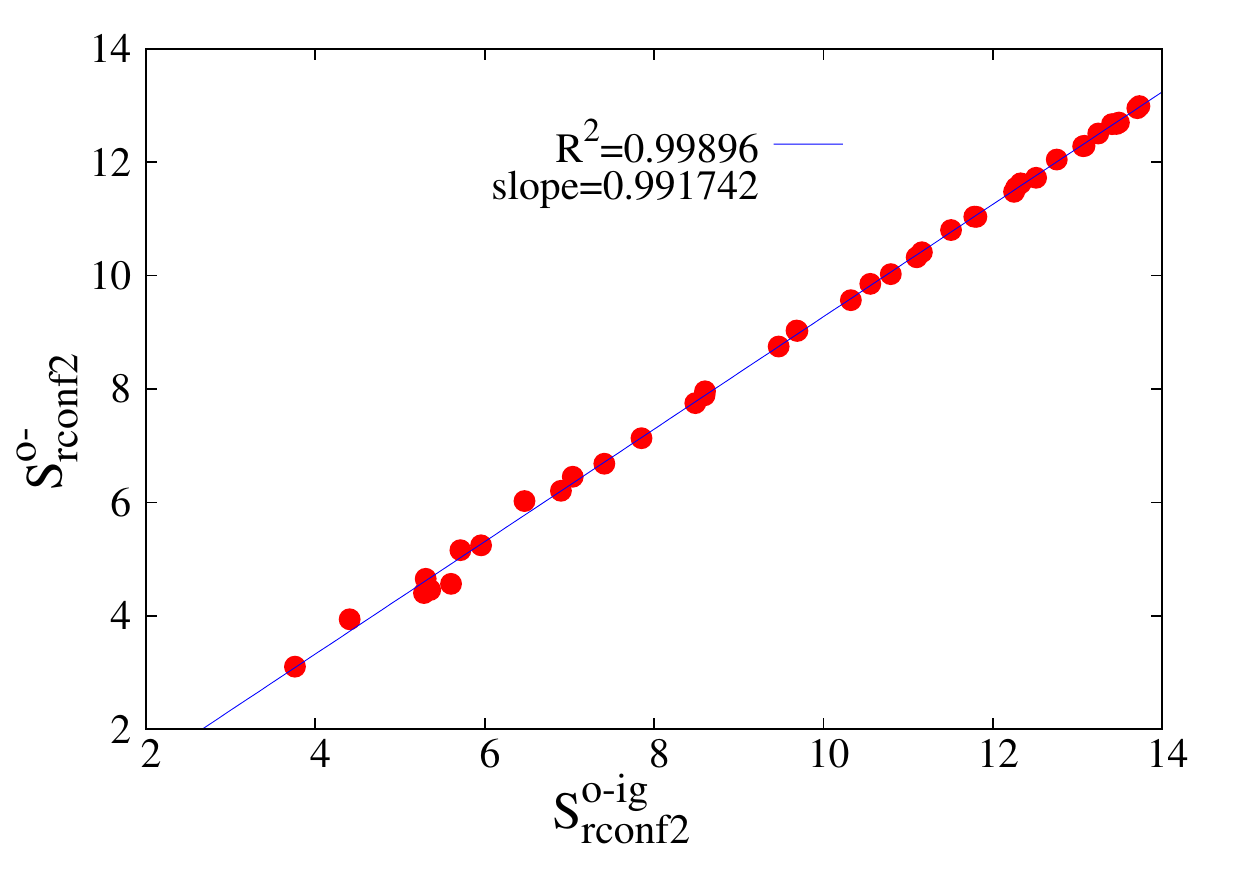}}\\
\subfloat[]{\includegraphics[width=3.in]{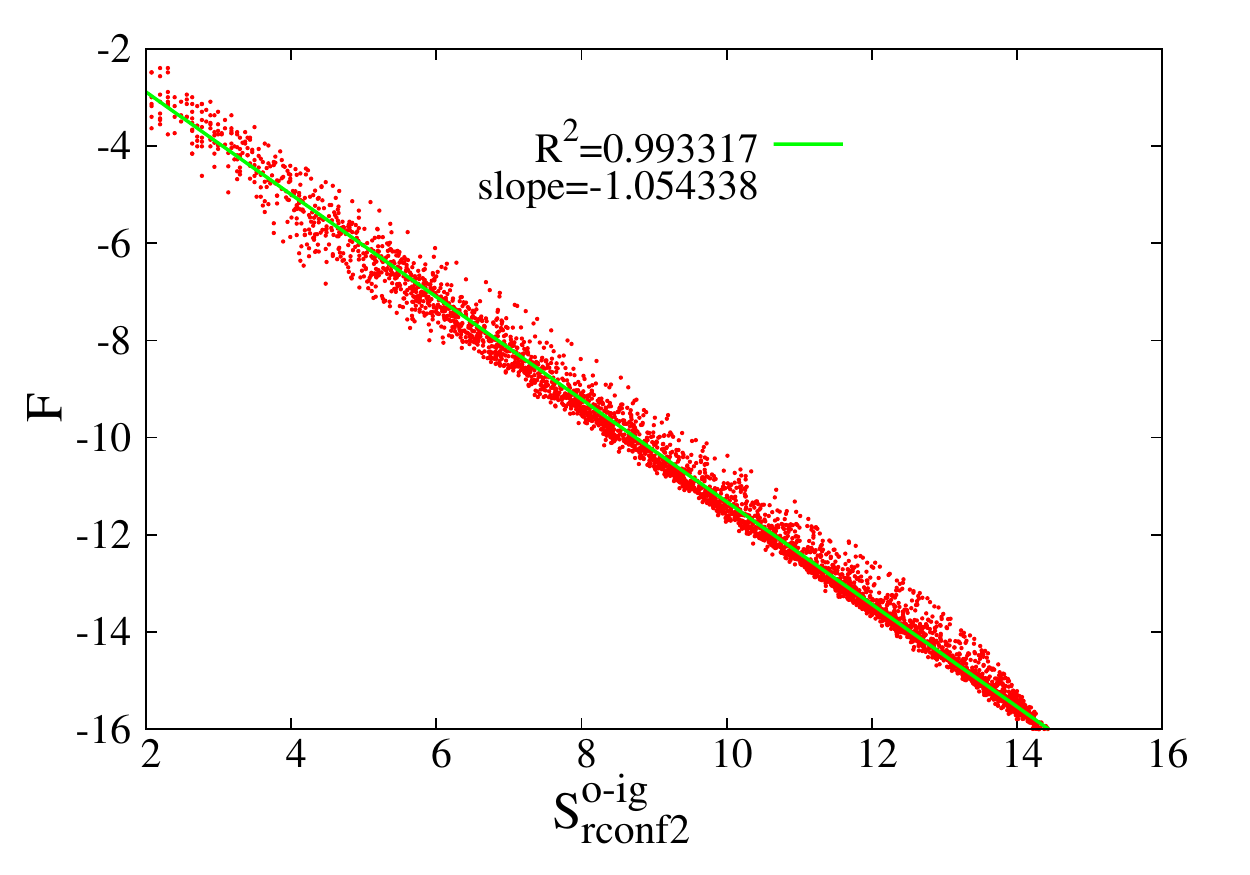}}
\subfloat[]{\includegraphics[width=3.in]{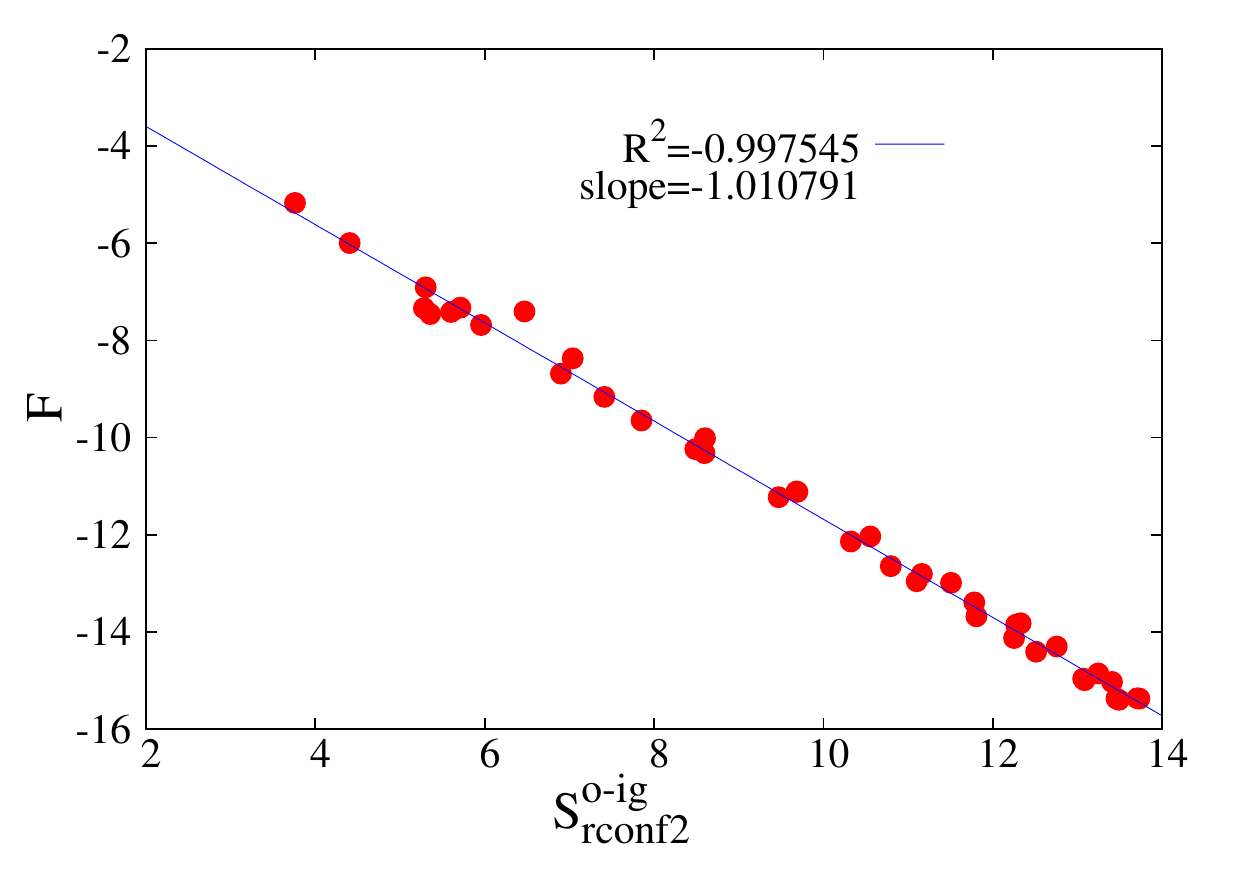}}
\caption{Results for RCONF2 based analysis of HEWL-b trajectory set. $S^{O-}_{rconf}$ vs. $S^{O-ig}_{rconf}$ (both in the unit of $k_B$) plots for a) 5120 partitions generated from projection onto 256 backbone dihedrals and b) 40 partitions generated from projection onto radius of gyration $R_g$ and the number of native contacts $N_{nc}$. c) and d) Relative free energy $F$ (in the unit of $k_BT$) vs. $S^{O-ig}_{rconf}$ (in the unit of $k_B$) plots for the same two sets of the conformational partitions.}
%\label{rconf2}
\end{figure}

%% graphical table of contents.
%%%%%%%%%%%%%%%%%%%%%%%%%%%%%%%%%%%%%%%%%%%%%%%%%%%%%%%%%%%%%%%%%%%%%
%\clearpage
%\begin{tocentry}
%\includegraphics[angle=-90,width=0.95\textwidth]{./Figures/Figure_content-eps-converted-to.pdf}
%\end{tocentry}

\end{document}